\documentclass{aa}  

\usepackage{graphicx}
\usepackage[dvipsnames]{xcolor}
\usepackage{txfonts}
\usepackage{booktabs}
\usepackage{bm}
\usepackage{ulem}
\newcommand{\dotprod}[0]{{\cdot}}
\newcommand{\mach}[0]{\mrm{Ma}}
\newcommand{\mrm}[1]{\mathrm{#1}}

\begin{document}

   \title{Towards a self-consistent model of the convective core boundary in upper main sequence stars. Part I: 2.5D and 3D simulations}
   \titlerunning{Towards a self-consistent model of the convective core boundary in upper main sequence stars}
   
   \author{
        R.~Andrassy\inst{1}\and
        G.~Leidi\inst{1}\and
        J.~Higl\inst{1,2}\and
        P.~V.~F.~Edelmann\inst{3}\and
        F.~R.~N.~Schneider\inst{1,4}\and
        F.~K.~R{\"o}pke\inst{1,5}
        }
    \institute{
        Heidelberger Institut f{\"u}r Theoretische Studien,
        Schloss-Wolfsbrunnenweg 35, D-69118 Heidelberg, Germany\\
        \email{robert.andrassy@h-its.org}
        \and
        High-Performance Computing Center Stuttgart,
        Nobelstraße 19, 70569 Stuttgart, Germany        
        \and
        Computer, Computational and Statistical Sciences (CCS) Division and Center for Theoretical
        Astrophysics (CTA), Los Alamos National Laboratory, Los Alamos, NM
        87545, USA
        \and
        Zentrum f{\"u}r Astronomie der Universit{\"a}t Heidelberg, Astronomisches Rechen-Institut, M{\"o}nchhofstr.\ 12-14, 69120 Heidelberg, Germany
        \and
        Zentrum f\"ur Astronomie der Universit\"at Heidelberg, Institut f\"ur
        Theoretische Astrophysik, Philosophenweg 12, D-69120 Heidelberg, Germany
   }
    
   \date{Received ; accepted }

  \abstract{There is strong observational evidence that the convective cores of intermediate-mass and massive main sequence stars are substantially larger than those predicted by standard stellar-evolution models. However, it is unclear what physical processes cause this phenomenon or how to predict the extent and stratification of stellar convective boundary layers. Convective penetration is a thermal-timescale process that is likely to be particularly relevant during the slow evolution on the main sequence. We use our low-Mach-number \textsc{Seven-League Hydro} code to study this process in 2.5D and 3D geometries. Starting with a chemically homogeneous model of a $15\,\mrm{M}_\odot$ zero-age main sequence star, we construct a series of simulations with the luminosity increased and opacity decreased by the same factor, ranging from $10^3$ to $10^6$. After reaching thermal equilibrium, all of our models show a clear penetration layer; its thickness becomes statistically constant in time and it is shown to converge upon grid refinement. The penetration layer becomes nearly adiabatic with a steep transition to a radiative stratification in simulations at the lower end of our luminosity range. This structure corresponds to the adiabatic `step overshoot' model often employed in stellar-evolution calculations. The simulations with the highest and lowest luminosity differ by less than a factor of two in the penetration distance. The high computational cost of 3D simulations makes our current 3D data set rather sparse. Depending on how we extrapolate the 3D data to the actual luminosity of the initial stellar model, we obtain penetration distances ranging from $0.09$ to $0.44$ pressure scale heights, which is broadly compatible with observations.}

   \keywords{hydrodynamics -- convection -- turbulence -- methods: numerical -- stars: massive -- stars: interiors}

   \maketitle
%

\section{Introduction}
\label{sec:introduction}

Numerous observations strongly suggest that the convective, hydrogen-burning cores of intermediate-mass and massive stars are larger than those predicted based on the linear-stability criteria of Schwarzschild and Ledoux \citep[see e.g.][]{kippenhahn2012a}. The evidence has traditionally been based on colour--magnitude diagrams of open clusters \citep[e.g.][]{maeder1981a, demarque1994a} and observations of eclipsing binary systems \citep[][and references therein]{claret2016a}. More recently, core sizes have also been measured using asteroseismology \citep[][and references in the latter]{aerts2013a, anders2023a}, confirming the large core radii. The size of the mixed core on the main sequence influences stellar lifetimes as well as stellar structure and evolution in later evolutionary stages.
The umbrella term `convective overshooting' has traditionally been used to describe all physical processes that contribute to extending stellar convection zones, although the terms `convective penetration' and `convective boundary mixing' are also used, often interchangeably. We reserve the term convective penetration for a process that is fast enough to change the thermal stratification beyond the Schwarzschild boundary \citep[see also][]{anders2023a}. Classical one-dimensional (1D) stellar evolution theory resorts to simple parametric prescriptions for such processes, which severely limits the predictive power of current stellar-evolution models. Two prescriptions are particularly popular \citep[see e.g.][]{kippenhahn2012a}: `step overshoot' and `exponentially decaying diffusion'. The first represents an approximate model of convective penetration. It assumes that mixing that is approximately as fast as that in the formally convective layer extends some distance beyond the Schwarzschild boundary, making the stratification chemically homogeneous and adiabatic with a discontinuous transition to the radiative stratification. The second prescription describes the mixing of chemical species using a diffusion coefficient decreasing exponentially with distance from the convective boundary while assuming that the thermal stratification remains radiative.\footnote{This prescription (see e.g.\ \citet{paxton2011a}, \citet{kippenhahn2012a}, or \citet{anders2023a} for details) is physically inconsistent close to the convection zone, where the chemical diffusion coefficient is often assumed to be much larger than the coefficient of radiative diffusion but the exchange of heat associated with the rapid exchange of species between fluid elements is ignored. Some authors also assume the temperature gradient to be radiative when using the step overshoot prescription (see \citet{anders2023a} for an overview).} Both prescriptions involve a free parameter that describes the radial extent of the mixing.

Core convection is characterised by low Mach numbers (of order $10^{-4}$ to $10^{-3}$) in main sequence stars. The overlying stratification is so stable that it stops the slow convective flows within a tiny fraction of the pressure scale height \citep{roxburgh1965b, saslaw1965a}. This kind of dynamical `overshoot' cannot reconcile stellar-evolution models with the observations mentioned above. Three-dimensional (3D), hydrodynamic simulations show that convective flows turning around at the convective boundary can entrain relatively small mass fractions of material originally located beyond the formal boundary of convective instability. This process of mass entrainment occurs even at very stiff convective boundaries \citep[e.g.][]{meakin2007a, woodward2015a, horst2021a}. However, simulations of core convection on the main sequence predict unrealistically high mass-entrainment rates \citep{meakin2007a, gilet2013a, herwig2023a}, suggesting that the observed fast entrainment is just a transient phenomenon. Indeed, most simulations of this kind do not include radiative diffusion, making it impossible to sustain the expected stellar structure with a convective core and a radiative envelope on long timescales. Particularly relevant is the thermal timescale on which heat transport processes, including convective penetration, set the thermal structure of the star.

Processes occurring on the thermal timescale can be qualitatively described in terms of how they affect the radial profile of entropy. Both nuclear burning and hydrodynamic entrainment of high-entropy material from the radiative envelope increase the mean entropy of the mixed core. On the other hand, radiative diffusion carries energy outwards, decreasing the core entropy. It is reasonable to assume that the two processes reach equilibrium and the core stops changing its size on the thermal timescale (ignoring slow changes in chemical composition occurring on the nuclear timescale). Somewhat surprisingly, analytical constraints can be placed on how large the convective core can be in the equilibrium state. \citet{roxburgh1978a, roxburgh1989a} averages the 3D equations of fluid motion and radiative heat transport in space and time and, assuming statistically stationary convection, derives a simple integral criterion for the size of the convective core:
\begin{align}
    \int_V \left( \overline{F} - \overline{\Gamma} \right) \frac{1}{T_0^2} \frac{\mrm{d} T_0}{\mrm{d} r} \mrm{d}V = \int_V \frac{\Phi_0}{T_0} \mrm{d}V, \label{eq:roxburgh_criterion}
\end{align}
where $\overline{F}$ and $\overline{\Gamma}$ are the mean radial components of the radiative flux and of the energy flux from nuclear sources, respectively, $T$ is the temperature, and $\Phi$ the dissipation rate of kinetic energy. The subscript `$0$' refers to the mean state and the integration is performed over volume $V$ of the core. The radial profile of the dissipation rate $\Phi_0$ depends on details of the turbulent convective flow. Neglecting this term, Roxburgh obtains the maximum possible core mass, concluding that it can be substantially larger than the mass inside the formal Schwarzschild boundary. \citet{roxburgh1978a, roxburgh1989a} as well as \citet{zahn1991a} argue that the whole core must be nearly adiabatic, including its extension beyond the Schwarzschild boundary now known as the `convective penetration layer'. However, the thickness of this layer cannot be further constrained without constraining the turbulent dissipation rate.

Multi-dimensional hydrodynamic simulations can encompass all physical processes needed to compute detailed models of penetrative convection. However, the thermal timescale associated with typical stellar convective layers is orders of magnitude longer than the convective turnover timescale and the only way to to obtain thermally relaxed simulations is to boost the heat flux by several orders of magnitude. Using this technique, a few research groups recently managed to approach the thermal timescale and to observe the formation of a convective penetration layer in various stellar environments \citep{hotta2017a, kapyla2019a, baraffe2023a, blouin2023a, mao2023a}. \citet{baraffe2023a} ran 2D simulations of a range of main sequence stars with convective cores but only one of their simulations (of a $3\,\mrm{M}_\odot$ star) uses sufficient luminosity boosting for the penetration layer to start approaching thermal equilibrium. The growth of the convective core of a $25\,\mrm{M}_\odot$ main sequence star slows down on the thermal timescale in the 3D simulations of \citet{mao2023a} but the core does not stop growing. This might be a consequence of continued chemical mixing in the simulation. In the simpler case of a chemically homogeneous stratification, \citet{anders2022a} show that the penetration layer stops growing on a sufficiently long timescale, reaching a statistically stationary state. However, their simulations use a simplified stratification rather than a realistic stellar model and their use of the Boussinesq approximation eliminates any effects of compressibility, which may be important in the strongly stratified stellar interior.

Our study focuses on thermal aspects of the stellar convective-penetration problem. We eliminate compositional effects by using a realistic model of a $15\,\mrm{M}_\odot$ zero-age main sequence (ZAMS) star. Employing the standard approach of luminosity boosting, we show that a penetration layer forms at the core boundary and stops growing on the thermal timescale. We measure the core size in the thermally relaxed state and extrapolate towards the star's nominal luminosity. We run both 2.5D (a sphere with assumed axial symmetry) and 3D simulations using our fully compressible, low-Mach number \textsc{Seven-League Hydro} (SLH) code \citep{miczek2013a, edelmann2014a, edelmann2021a} to explore both the limit of low turbulent dissipation, as expected in 2D convection and assumed by Roxburgh, and the more realistic case of 3D turbulent convection. 

We describe our 1D initial stellar model in Sections~\ref{sec:stellar-model} and \ref{sec:stellar-model-simplifications}. The numerical setup of the 2.5D and 3D SLH simulations is detailed in Sect.~\ref{sec:hydro-simulations}. We use a simple numerical experiment to illustrate the importance of radiative diffusion in Sect.~\ref{sec:importance-of-diffusion}. Relevant properties of our 2.5D and 3D simulations and their numerical convergence are described in Sections~\ref{sec:velocity-field}--\ref{sec:numerical-convergence}. We then extrapolate the thickness of the penetration layer to the actual luminosity of the stellar model in Sect.~\ref{sec:luminosity-dependence}. Our results are summarised and discussed in Sect.~\ref{sec:summary-discussion}.

\section{Methods}
\label{sec:methods}
\subsection{One-dimensional stellar model}
\label{sec:stellar-model}

We use version $15140$ of the stellar-evolution code MESA \citep{paxton2011a, paxton2013a, paxton2015a, paxton2018a, paxton2019a, jermyn2023a} to compute a model of a $15\,\mrm{M}_\odot$ star with a metallicity of $Z = 0.02$ at the ZAMS. The evolution model is stopped at the age of $7.96 \times 10^4$\,yr, when the convective core has fully developed and the star has reached thermal equilibrium. At this stage, the central hydrogen mass fraction has decreased by $0.156\%$ from its initial value of $0.700$. The star's radius and luminosity are \mbox{$R_\star = 3.47 \times 10^{11}$\,cm} ($4.98\,\mrm{R}_\odot$) and $L_\star = 7.49 \times 10^{37}$\,erg\,s$^{-1}$ ($1.95 \times 10^4\,\mrm{L}_\odot$), respectively. The model does not include any convective `overshoot' or penetration. The latter evolves in our multi-dimensional models in a self-consistent way. The Schwarzschild boundary of the core is located at $8.74 \times 10^{10}$\,cm ($0.252\,R_\star$). The MESA inlist for the model as well as the resulting profile file are available on Zenodo\footnote{\url{https://doi.org/10.5281/zenodo.8127093}}.

\subsection{Simplification of the initial 1D model}
\label{sec:stellar-model-simplifications}

\begin{figure}
\includegraphics[width=\linewidth]{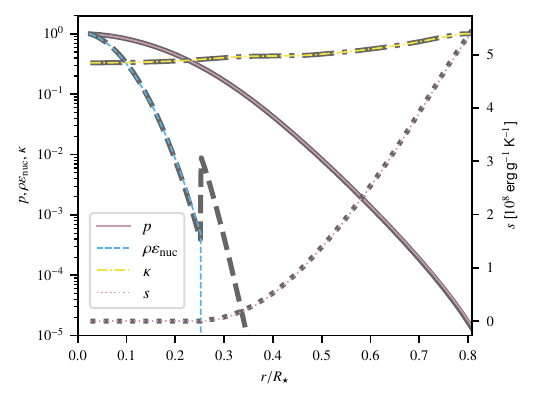}
\caption{Stratification of the pressure $p$, energy generation rate per unit volume $\rho\varepsilon_\mrm{nuc}$, opacity $\kappa$, and specific entropy $s$ in the MESA model (thick grey lines) and in the SLH model at $t = 0$ (thin coloured lines). The profiles of $p$ and $\rho\varepsilon_\mrm{nuc}$ are normalised such that their maximum values are unity. The entropy of the convective core is set to $0$. The jump in $\rho \varepsilon_\mrm{nuc}$ at the boundary of the fully-mixed core is caused by a slight change in chemical composition, which is neglected in the SLH model.}
\label{fig:initial_stratification}
\end{figure}

The MESA model contains a $0.093\%$ jump in the mean molecular weight at the core boundary because some hydrogen burning occurred during the initial thermal adjustment of the model. We remove this step by applying the central value of the mean molecular weight $\mu = 0.6174$ to the whole star in order to eliminate any compositional effects. To obtain an initial condition for multidimensional simulations, we re-integrate the hydrostatic stratification assuming that the difference between the actual temperature gradient $\nabla$ and the adiabatic gradient $\nabla_\mrm{ad}$ is given by the MESA model everywhere but in the convective core, where it is set to zero. The equation of state used in this integration process as well as in the hydrodynamic simulations described below accounts for a mixture of fully ionised, monatomic gas and radiation. The radial profile of gravitational acceleration is interpolated from the MESA model without any modification. The resulting profiles of the pressure $p$ as well as specific entropy $s$ closely match those of the original MESA model, see Fig.~\ref{fig:initial_stratification}.

Figure~\ref{fig:initial_stratification} also shows the profiles of opacity $\kappa$ and energy generation rate per unit volume $\rho \varepsilon_\mrm{nuc}$. The opacity is simply interpolated without any modification. The profile of the energy generation rate is for reasons of computational efficiency modelled using the fitting function
\begin{equation}
    \rho \varepsilon_\mrm{nuc} = 
    \begin{cases}
        4.135 \times 10^5 \exp\left[-\frac{r^2}{(3.15 \times 10^{10}\,\mathrm{cm})^2}\right]\,\mrm{erg\,cm}^{-3}\ \mrm{for}\ r < r_0,\\
        0\ \mrm{otherwise},
    \end{cases}
\end{equation}
where $r_0 = 8.736 \times 10^{10}\,\mrm{cm}$ is the radius of the initial core boundary. This profile is almost indistinguishable from that provided by MESA for $r < r_0$, see Fig.~\ref{fig:initial_stratification}. There is a small-amplitude spike in the energy generation rate at $r = r_0$ in the MESA model due to a slight change in composition. We do not model this spike. Additionally, our 2.5D and 3D simulations have a central cut-out with a radius of $8.7 \times 10^9$\,cm ($0.025\,R_\star$, see Sect.~\ref{sec:hydro-simulations} for details). Taking into account all of these effects, the total luminosity of the modified model is $5.5\%$ lower than that of the original MESA model.

\subsection{2.5D and 3D simulations}
\label{sec:hydro-simulations}

As mentioned in Sect.~\ref{sec:introduction}, the time and length scales involved in the process of convective penetration in stars make the problem prohibitively costly for multidimensional hydrodynamic simulations. We illustrate this in Table~\ref{tab:scales}, which is based on results of our simulations (see Sect.~\ref{sec:results} for details). The last row of the table is an extrapolation to our $15\,\mrm{M}_\odot$ star's actual parameters. The thermal timescale of the model is $5.9 \times 10^4$\,yr, which corresponds to $6.5 \times 10^5$ convective turnover timescales. This is well beyond what can currently be achieved using multidimensional simulations. Additionally, we must consider the need to resolve the length scale of radiative diffusion on the turnover timescale. Table~\ref{tab:scales} shows that this length scale evaluated at the core boundary is only $0.4$ cell widths on a grid with $512$ radial grid cells. To marginally resolve it by $4$ cells, we would need a grid with ${\approx}\,5000$ radial cells, making the simulations extremely expensive. 

We solve these problems by increasing the energy-generation rate by a boost factor $b$ ranging from $10^3$ to $10^6$, which shortens both the thermal and convective turnover timescales. We decrease the opacity by the same factor, which increases the thermal diffusion length scale. This choice also has the advantage that the radiative temperature gradient $\nabla_\mrm{rad}$, which is proportional to the product of the luminosity and opacity, becomes independent of $b$ at a given temperature and density. The stratification of the radiative layer remains largely the same, although slight changes ($\lesssim 25\%$ in $\nabla / \nabla_\mrm{ad}$) still occur near the outer boundary of the simulation box ($r / R_\star \gtrsim 0.7$) as the hydrostatic stratification adjusts in response to (1) an initial growth of the convective core (see Sect.~\ref{sec:thermal-evolution} for details) and (2) the decreased luminosity as compared with the MESA model (by $5.5\%$, see Sect.~\ref{sec:stellar-model-simplifications} for details).

\begin{table}
\caption{Time and length scales for the $15\,\mrm{M}_\odot$ ZAMS star.}
\label{tab:scales}
\centering
\begin{tabular}{lllll}
\toprule
$b$ & $\tau_\mrm{th}$ [yr] & $\tau_\mrm{conv}$ [d] & $\tau_\mrm{th} / \tau_\mrm{conv}$ & $l_\mrm{d} / \Delta r_{512}$ \\
\midrule
$10^6$ & $5.9 \times 10^{-2}$ & $7.5 \times 10^{-1}$ & $2.9 \times 10^1$ & $69$ \\
$10^5$ & $5.9 \times 10^{-1}$ & $1.4 \times 10^0$ & $1.5 \times 10^2$ & $29$ \\
$10^4$ & $5.9 \times 10^0$ & $2.6 \times 10^0$ & $8.1 \times 10^2$ & $12$ \\
$10^3$ & $5.9 \times 10^1$ & $5.0 \times 10^0$ & $4.3 \times 10^3$ & $5.0$ \\
\midrule
$10^0$ & $5.9 \times 10^4$ & $3.3 \times 10^1$ & $6.5 \times 10^5$ & $0.4$ \\
\bottomrule
\end{tabular}
\vspace{0.5em}
\tablefoot{The variables in the header are the boost factor $b$, thermal timescale $\tau_\mrm{th}$, convective turnover timescale $\tau_\mrm{conv}$, radiative diffusion length scale $l_\mrm{d}$ on the convective turnover timescale, and the radial grid spacing $\Delta r_{512}$ in a simulation with $512$ radial cells.}
\end{table}

We use the SLH code to simulate the convective penetration process using the simplified 1D model as an initial condition. The code solves the following set of compressible, inviscid Euler equations with gravity and diffusive heat transport:
\begin{align}
   \partial_t\, \rho + \bm{\nabla} \dotprod (\rho \bm{u}) &= 0,
   \label{eq:euler_rho} \\ \partial_t (\rho \bm{u}) + \bm{\nabla} \dotprod (\rho
   \bm{u} \otimes \bm{u} + p\bm{\mathbb{I}}) &= \rho\bm{g},
   \label{eq:euler_rhou} \\
   \partial_t (\rho e_\mrm{t}) + \bm{\nabla} \dotprod [(\rho e_\mrm{t} +
   p)\bm{u} - \chi \bm{\nabla} T] &= \rho \varepsilon_\mrm{nuc}, \label{eq:euler_rhoe}
\end{align}
where $\rho$, $\bm{u}$, $p$, $\bm{\mathbb{I}}$, $\bm{g}$, $\chi$, $T$, $\varepsilon_\mrm{nuc}$ denote
the density, velocity, pressure, unit tensor, gravity, thermal conductivity,
temperature, and energy generation rate by nuclear reactions, respectively. The specific total energy
$e_\mrm{t} = e_\mrm{i} + e_\mrm{k} + \Psi$ includes internal energy $e_\mrm{i}(\rho, T)$,
kinetic energy $e_\mrm{k} = \frac{1}{2} |\bm{u}|^2$, and a time-independent
gravitational potential $\Psi$.

The equations are solved on two types of grids: 2.5D and 3D. Both of them match the spherical geometry of stars to suppress discretisation errors. The 3D grid is a spherical grid with uniform spacing in the radius $r$, polar angle (colatitude) $\vartheta$, and azimuthal angle (longitude) $\varphi$. The 2.5D grid is a polar grid with uniform spacing in $r$ and $\vartheta$ that describes a 3D spherical system with all variables constant in $\varphi$ (rotational symmetry around the polar axis). This results in a geometric source term in the numerical scheme, which is taken into account. The grids cover the radial range from $8.7 \times 10^9$\,cm ($0.025\,R_\star$) to $2.8 \times 10^{11}$\,cm ($0.808\,R_\star$). We limit the range of polar angles to $30^\circ \le \vartheta \le 150^\circ$ because the cell width in the azimuthal direction drops close to the polar axis, severely limiting the time-step length. In the 3D case, we include azimuthal angles in the range $-60^\circ \le \varphi \le 60^\circ$.

To judge numerical convergence of our results, we run simulations with each boost factor $b$ on a range of grids. The number of radial grid cells is $N_r \in (128, 256, 512, 1024)$ in 2.5D and $N_r \in (128, 192, 256)$ in 3D. The only exception is that we do not include a 2.5D simulation with $b = 10^3$ and $N_r = 1024$, which would be too costly. The number of cells in each of the two angular directions is always $N_r/2$.

We impose reflective boundary conditions \citep[see e.g.][]{leveque2002a} at all boundaries of the computational domain. This is implemented via ghost cells filled such that the component of momentum normal to the boundary becomes an odd function and all other variables become even functions of the distance from the boundary.\footnote{The even symmetry implies that the boundary does not restrict the component of momentum parallel to the boundary, i.e. we have a free-slip boundary.} The only exception is the outer radial boundary, where the temperature in the first ghost cell is set to the value given by the initial 1D model when and only when the radiative-diffusion term is computed.\footnote{The alternative of keeping the outgoing radiative flux constant becomes unstable when the opacity is assumed constant as is the case in our simulations. A large-enough negative temperature fluctuation at the boundary strongly reduces thermal conductivity $\chi \propto T^3$, blocking the flux from deeper layers. The constant-flux boundary condition makes the temperature drop further, closing a positive feedback loop, which ultimately reduces the temperature to zero close to the boundary.} The reflective boundaries in the angular directions eliminate horizontal shear flows, which often become strong with periodic boundaries.

The SLH code is based on a finite-volume discretisation and offers both explicit and implicit time steppers via the method of lines. In this work, we use the ESDIRK23 implicit stepper \citep{hosea1996a}. The bulk Mach numbers reported in Sect.~\ref{sec:velocity-field} may suggest that the flows in some of our simulations are fast enough to be better suited for explicit time steppers. However, this impression is misleading. Even without considering radiative diffusion, the length of explicit time steps would be limited by the inner edge of the computational grid, where the sound speed is fastest and grid cells are narrowest in the angular directions. For this reason, implicit time steps can be $50$ to $80$ times longer than explicit ones even with the highest boost factor $b = 10^6$. This ratio is of similar order as the cost-per-step ratio of the implicit to a comparable explicit stepper, making both approaches comparably expensive with $b = 10^6$. However, fast radiative diffusion caused by low density at the outer edge of the simulation domain imposes much stricter time-step constraints. The ratio of the lengths of implicit to diffusion-limited explicit time steps in the 2.5D simulation with $b = 10^6$ and computed on the $1024 \times 512$ grid is ${\approx}\,5000$. Diffusion is much less constraining in simulations with $b = 10^3$ but convection is so slow in those that than longest possible explicit time steps would be ${\approx}\,500$ times shorter than implicit steps. All in all, the implicit approach saves a significant amount of computing time.\footnote{In principle, we could treat only the diffusion term implicitly in simulations with high boost factors but this feature has not been implemented in SLH.}

We use unlimited parabolic reconstruction in space. Because the stratification spans five orders of magnitude in pressure (see Fig.~\ref{fig:initial_stratification}), hydrostatic equilibrium requires a special treatment to suppress spurious flows caused by discretisation errors, see \citet{edelmann2021a} for details. Specifically, we use the well-balancing method of \citet{berberich2021a}, later referred to as the `deviation method' by \citet{edelmann2021a}. We further increase our resolving power by employing the low-Mach-number numerical flux function AUSM+-up of \citet{liou2006a}, which is much less dissipative than standard Riemann solvers in slow flows. We slightly modify the flux function as described by \citet{edelmann2021a}. The modified flux uses the parameters $f_a = 10^{-10}$ and $f_a^p = 10^{-1}$ in the notation of the latter paper. 

Although our numerical solver is deterministic, turbulent convection is well known to be highly sensitive to initial conditions and perturbations. A pair of simulations with initial conditions differing by small amounts or one with identical initial conditions but a small change to the implicit time stepper (e.g.\ allowing one more iteration per time step) are likely to produce completely different flow patterns after a few convective timescales. Time-averaged quantities become well defined when averaging over many convective timescales, assuming that there is a statistically stationary state. However, the statistical nature of turbulence must still be taken into account when comparing averages from different simulations. We take great care to quantify statistical variation our simulation data. We express statistical-variation ranges using $\pm 1 \sigma$ intervals, where $\sigma$ is the uncorrected standard deviation of the statistical sample, unless mentioned otherwise.

\section{Results}
\label{sec:results}

\subsection{Importance of radiative diffusion}
\label{sec:importance-of-diffusion}

\begin{figure}
\includegraphics[width=\linewidth]{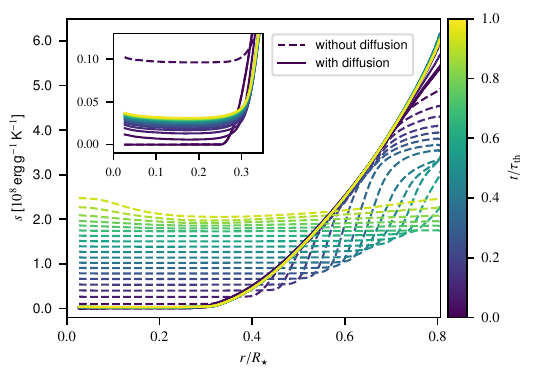}
\caption{Time evolution of the radial profiles of the specific entropy $s$ in 2.5D simulations without and with radiative diffusion. Both simulations have $b = 10^4$ and $N_r = 256$. The profiles are shifted such that the entropy of the convective core is $0$ at $t = 0$. Simulation time is given as a fraction of the thermal timescale $\tau_\mrm{th}$ on the colour bar. Each curve, except for that showing the initial condition, corresponds to a time average over $0.05 \tau_\mrm{th}$.}
\label{fig:with-vs-without-diffusion}
\end{figure}

We start by demonstrating that the presence of radiative diffusion is essential for our simulations to reach a statistically stationary state. In Fig.~\ref{fig:with-vs-without-diffusion}, we compare two simulations performed with a boost factor of $10^4$ on a $256 \times 128$ grid. The convective core keeps growing in the simulation that does not include radiative diffusion until, ultimately, the whole model becomes convective. This effect can be understood even without considering hydrodynamic mass entrainment. The heat source in the core keeps increasing the mean entropy of the mixed core. Whenever that entropy becomes greater than the entropy of a stably stratified layer atop the core, that layer gets mixed into the core. Ultimately, the entropy of the whole model becomes approximately the same and the star becomes fully convective. This process is further accelerated by the presence of strong internal gravity waves, which seem to induce some mixing and flatten the entropy profile close to the outer boundary of the computational domain (i.e.\ where the waves reach their maximum amplitude). On the other hand, changes to the entropy profile are much more subtle in the simulation with radiative diffusion, in which the initial outward propagation of the convective boundary stops early on and the simulation reaches a statistically stationary state. This is possible because radiative diffusion is sufficiently fast to transport entropy generated in the core through the radiative envelope. The heat then leaves the box owing to the constant-temperature outer boundary condition. This statistically stationary state can be maintained indefinitely because we do not model changes to the chemical composition of the core due to nuclear burning. We only discuss simulations with radiative diffusion in the rest of this section.

\subsection{Velocity field}
\label{sec:velocity-field}

\begin{figure*}
\includegraphics[width=\linewidth]{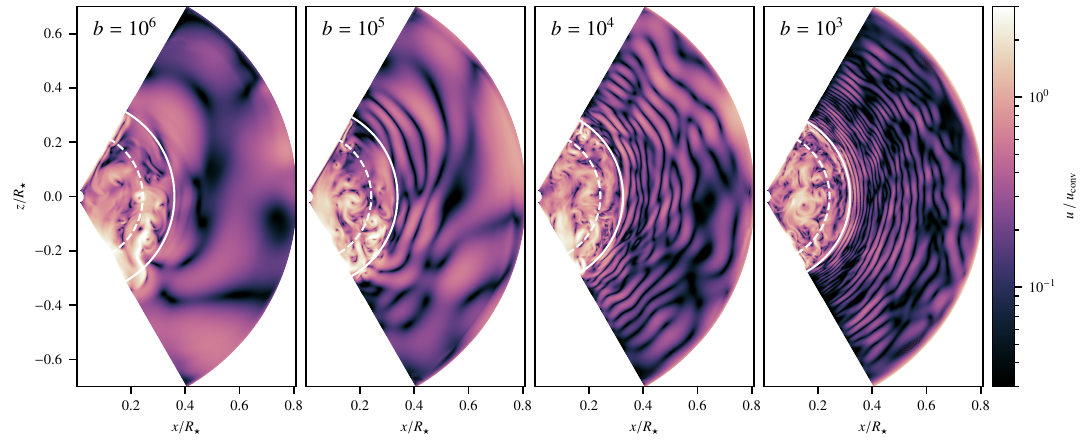}
\caption{Velocity fields in 2.5D simulations with boost factors $b \in (10^6, 10^5, 10^4, 10^3)$ performed on the $512 \times 256$ grid. The velocity is normalised using the typical convective velocity $u_\mrm{conv}$ as given by the scaling law shown in Fig.~\ref{fig:u_conv_vs_b}. The dashed and solid lines give the time-averaged radii of the Schwarzschild and convective boundaries, respectively. The simulations are shown at $t = 0.5 \tau_\mrm{th}(b)$.}
\label{fig:velocity-distribution-2.5D}
\end{figure*}

\begin{figure}
\includegraphics[width=\linewidth]{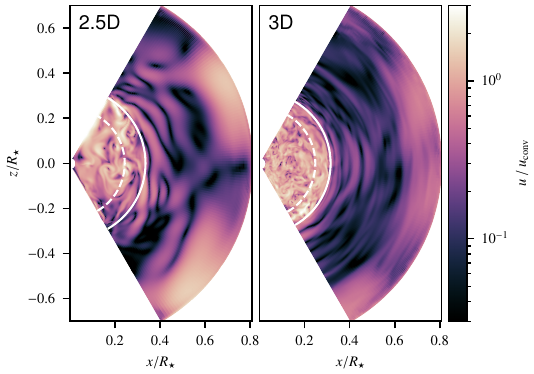}
\caption{As in Fig.~\ref{fig:velocity-distribution-2.5D}, but comparing 2.5D and 3D simulations with a boost factor of $b = 10^5$ performed on grids of $256 \times 128$ and $256 \times 128^2$ cells, respectively. In the 3D case, a slice with the spherical angle $\varphi =0$ is shown.}
\label{fig:2.5D-vs-3D}
\end{figure}

The velocity field depends primarily on the boost factor $b$. This factor sets both convective velocity and efficiency. Because we scale $\kappa \propto b^{-1}$, the rate of radiative diffusion increases in proportion to $b$ and fluid parcels exchange increasing amounts of heat with their surroundings as they rise and sink in the convective core. This effect makes convection less efficient at transporting heat. Figure~\ref{fig:velocity-distribution-2.5D} shows the velocity field in 2.5D simulations with boost factors ranging from $10^6$ down to $10^3$. The velocity field in the core is dominated by large-scale vortices as typical of 2D convection. Small-scale motions are further suppressed by radiative damping. The structure of the convective flow is completely different in 3D, see Fig.~\ref{fig:2.5D-vs-3D}. Instead of large vortices, we obtain a turbulent cascade from large to small scales as expected for 3D convection (without strong rotation or magnetic fields).

The convective core generates internal gravity waves (IGWs), which propagate through the radiative envelope. When we decrease the boost factor $b$, the convection becomes slower and it generates waves at lower temporal frequencies. Basic theory of linear IGWs \citep[see e.g.][]{lighthill2001a, sutherland2010a} predicts that a decrease in frequency corresponds to a decrease in the radial wavelength for the same horizontal wavelength. This effect is evident in Fig.~\ref{fig:velocity-distribution-2.5D} and it is further amplified by the dependence on $b$ of the radiative diffusivity, which filters out the shortest wavelengths.

\begin{figure}
\includegraphics[width=\linewidth]{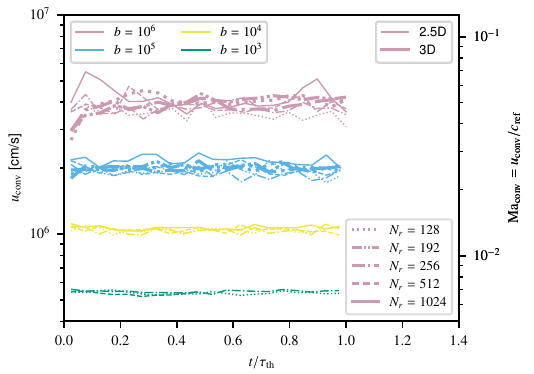}
\caption{Time evolution of the convective velocity $u_\mrm{conv}$ in all of our simulations with radiative diffusion. The time axis shows simulation time as a fraction of the thermal timescale $\tau_\mrm{th}(b)$. Each data point corresponds to a time average over $0.05 \tau_\mrm{th}$. The right vertical axis shows the corresponding Mach numbers computed using the reference sound speed $c_\mrm{ref} = 7.96 \times 10^7$\,cm\,s$^{-1}$.}
\label{fig:u_conv_evolution}
\end{figure}

The simulations in Fig.~\ref{fig:velocity-distribution-2.5D} are shown after a statistically stationary state has been reached. The figure also shows the radius of the convective boundary $r_\mathrm{cb}$ (solid lines), the exact definition of which we defer to the next section. For now, we only use it to measure convective velocity $u_\mrm{conv}$. We define this quantity to be the the mass-weighted, root-mean-square (rms) velocity averaged between the inner boundary of the computational domain and the radius $r_\mathrm{cb}$ of the convective boundary. Figure~\ref{fig:u_conv_evolution} shows $u_\mrm{conv}$ as a function of time, boost factor, grid resolution, and the dimensionality of the simulation. The convective velocity rapidly reaches a statistically stationary state. For each boost factor $b$, the velocity is not only statistically constant in time but it is also the same for 2.5D and 3D simulations and all grid resolutions. This is rather surprising since the structure of the convective flows differs substantially between 2.5D and 3D (Fig.~\ref{fig:2.5D-vs-3D}). Although the same total energy flux has to be transported in both cases, the velocity is only one of several variables that contribute to the fluxes of enthalpy and kinetic energy. Indeed, numerical studies of stellar convection usually show faster convective velocities in 2D than in 3D with the same initial condition \citep{muthsam1995a, meakin2007a, pratt2020a, horst2021a}. It is possible that the growth of convective instability is in our simulations limited by the same phenomenon in both geometries, which could be radiative damping or buoyancy braking in the penetration layer.

To shed more light on this issue, we show in Appendix~\ref{sec:core_convection_experiment} a set of 2.5D simulations with $b = 10^4$ that include radiative diffusion but eliminate mass entrainment from the stably stratified envelope. These simulations develop smooth, large-scale convective cells much faster than those in the analogous simulations of convective penetration. Moreover, the velocity in the large cells does not converge upon grid refinement, which is understandable considering the fact that numerical viscosity is the only mechanism dissipating kinetic energy in these simulations. On the other hand, the convective velocity is numerically converged in the simulations of convective penetration, suggesting that the growth of kinetic energy in those simulations is limited by a resolved, physical effect. We suspect that the buoyancy of the high-entropy material entrained from the radiative envelope into the core plays an important role in determining the convective velocity.

\begin{figure}
\includegraphics[width=\linewidth]{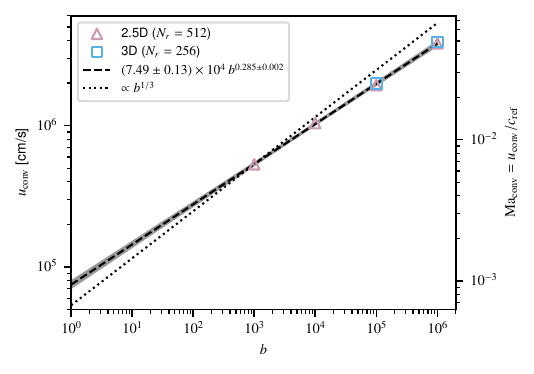}
\caption{Dependence of the convective velocity $u_\mrm{conv}$ on the boost factor $b$.  The right vertical axis shows the corresponding Mach numbers computed using the reference sound speed $c_\mrm{ref} = 7.96 \times 10^7$\,cm\,s$^{-1}$. Statistical-variation ranges for the individual data points are smaller than the markers and not shown. A straight line is fitted to the 2.5D data points using the Monte-Carlo procedure described in Sect.~\ref{sec:velocity-field}. The grey shading shows the $3\sigma$ statistical-variation interval around the mean fit. Parameters of the scaling law are given in the legend. The usual adiabatic scaling $u_\mrm{conv} \propto b^{1/3}$ is also shown for comparison.}
\label{fig:u_conv_vs_b}
\end{figure}

Convective velocity scales with $b^{1/3}$ in simulations of adiabatic convection \citep{jones2017a, cristini2019a, edelmann2019a, horst2021a, herwig2023a}. Our best-fit scaling is $u_\mrm{conv} \propto b^{0.285 \pm 0.002}$ for the four 2.5D simulations with $N_r = 512$, see Fig.~\ref{fig:u_conv_vs_b}. The mean exponent of the scaling differs by $-26\sigma$ from the adiabatic value of $1/3$. To obtain a measure of statistical variation for each data point, we consider the convective velocities averaged over each of the $n_\mrm{bins} = 14$ time bins after the initial transient ($t > 0.3\tau_\mrm{th}$) as statistically independent measurements. The standard deviation of the average of all of the bins is then estimated to be $\sigma = \sigma_\mrm{series} / n_\mrm{bins}^{1/2}$, where $\sigma_\mrm{series}$ is the standard deviation of the time series. We then compute $10^5$ statistical realisations of the scaling law assuming a normal distribution of statistical fluctuations, fit a power law to each of them, and compute the mean scaling and the statistical spread around the mean as a function of $b$. All 2.5D and 3D data points shown in Fig.~\ref{fig:u_conv_vs_b} are statistically consistent with the scaling law.\footnote{The most significant deviation is $1.7\sigma$ for the 3D simulation with $b = 10^6$.} Our best estimate of convective velocity at the star's nominal luminosity (i.e.\ $b = 1$) is $(7.49 \pm 0.13) \times 10^4$\,cm\,s$^{-1}$. Taking the mass-weighted average sound speed $c_\mrm{ref} = 7.96 \times 10^7$\,cm\,s$^{-1}$ inside the initial Schwarzschild radius as a reference, we obtain a nominal Mach number of $(9.41 \pm 0.16) \times 10^{-4}$. If we only fit the two lowest-luminosity 2.5D data points the resulting scaling exponent is $0.291 \pm 0.004$, which is statistically consistent with the exponent based on the full 2.5D data set. Finally, if we only fit the two 3D data points we obtain a scaling exponent of $0.295 \pm 0.006$, which differs by $-6.3 \sigma$ from the adiabatic value of $1/3$ and is statistically consistent with the exponent based on the full 2.5D data set. The 3D-based scaling gives a convective velocity of $(6.65 \pm 0.50) \times 10^4$\,cm\,s$^{-1}$ at $b = 1$, which is consistent with the 2.5D-based prediction. All of the 2.5D data points are also statistically consistent with the 3D-based scaling.\footnote{The most significant deviation is $1.6\sigma$ for the 2.5D simulation with $b = 10^5$.} It is possible that radiative damping makes the scaling shallower, although it is unclear why its exponent does not approach $1/3$ at low boost factors, i.e. when radiative diffusion becomes relatively slow in the core. One might also suspect numerical effects. However, \citet{edelmann2021a} demonstrate that the SLH code produces the expected $1/3$ velocity scaling for adiabatic convection down to Mach numbers of $2 \times 10^{-4}$ in a star-like environment. Simulations of 3D turbulent convection with mass entrainment performed using the SLH code have also been shown to closely agree with those obtained using another four major hydrodynamic codes used in the field, see \citet{andrassy2022a}. Finally, we provide a set of test simulations in Appendix~\ref{sec:code_accuracy}, which show that the results of SLH simulations converge upon grid refinement.

We define the convective turnover timescale as
\begin{align}
    \tau_\mrm{conv} = \frac{2 r_\mrm{cb}}{u_\mrm{conv}}.
\end{align}
For simplicity, we set $r_\mrm{cb} = r_{\mrm{Sb},0} + \alpha(b) H_{\mrm{p,Sb},0}$ in this estimate, where the $r_{\mrm{Sb},0}$ and $H_{\mrm{p,Sb},0}$ are the radius of and pressure scale height at the Schwarzschild boundary in the initial MESA model, and we take the scaling law $\alpha(b)$ derived from our 2.5D simulations in Sect.~\ref{sec:luminosity-dependence} and shown in Fig.~\ref{fig:alpha_vs_b_2}. The resulting values of $\tau_\mrm{conv}$, summarised in Table~\ref{tab:scales}, range from $0.75$\,d at $b = 10^6$ to $5.0$\,d at $b = 10^3$. Extrapolating to the nominal luminosity ($b = 1$), we obtain $\tau_\mrm{conv} = 33$\,d.

\subsection{Thermal evolution}
\label{sec:thermal-evolution}

\begin{figure}
\includegraphics[width=\linewidth]{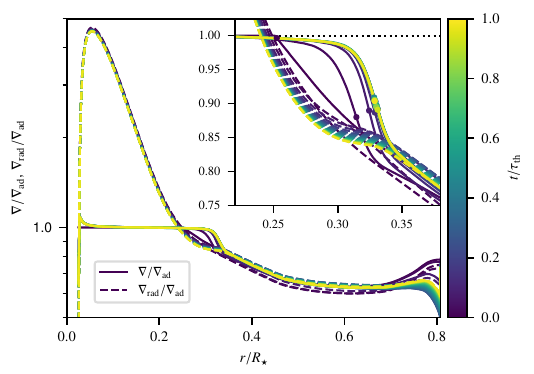}
\caption{Time evolution of the radial profiles of the actual and radiative temperature gradients $\nabla$ and $\nabla_\mathrm{rad}$, respectively, in the 2.5D simulation with $b = 10^4$ and $N_r = 1024$. The gradients are normalised using the adiabatic temperature gradient $\nabla_\mrm{ad}$. Simulation time is given as a fraction of the thermal timescale $\tau_\mrm{th}$ on the colour bar. Each curve, except for that showing the initial condition, corresponds to a time average over $0.05 \tau_\mrm{th}$. The inset zooms in onto the convective-penetration layer with dots marking the location of the steepest gradient in $\nabla / \nabla_\mrm{ad}$.}
\label{fig:F71-nabla-evolution}
\end{figure}

In all of our simulations, we initially observe rapid growth in the size of the convective core. To illustrate this, we plot in Fig.~\ref{fig:F71-nabla-evolution} the time evolution of the actual and radiative temperature gradients $\nabla$ and $\nabla_\mathrm{rad}$ in the 2.5D simulation with $b = 10^4$ and $N_r = 1024$. The gradients are normalised using the adiabatic temperature gradient $\nabla_\mrm{ad}$, so the formal Schwarzschild boundary is located at the radius $r_\mrm{Sb}$ where $\nabla_\mrm{rad} / \nabla_\mrm{ad} = 1$. The Schwarzschild boundary moves slightly inwards over the thermal timescale. In contrast, the growth in the radial extent of the nearly-adiabatic core is much more pronounced. A convective-penetration layer develops above the Schwarzschild boundary. The temperature gradient in the bulk of this layer is slightly sub-adiabatic but significantly super-radiative, i.e. the total flux could in principle be transported by radiative diffusion but vigorous convection is still present. The outward propagation rate of the convective boundary drops rapidly after ${\approx}\,0.1 \tau_\mrm{th}$ and a statistically stationary state seems to have been reached by ${\approx}\,(0.2-0.3) \tau_\mrm{th}$. There is some slight late-time adjustment of the temperature gradient near the outer boundary of the simulation box ($r/R_\star \gtrsim 0.7$). However, the layer affected by it is sufficiently far out to not influence the penetration distance.

\begin{figure}
\includegraphics[width=\linewidth]{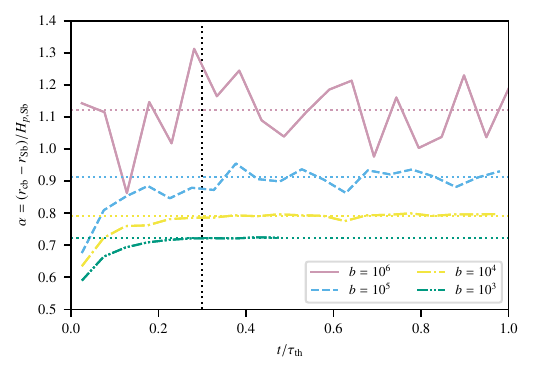}
\caption{Time evolution of the dimensionless penetration distance \mbox{$\alpha = (r_\mrm{cb} - r_\mrm{Sb})/H_{p,\mrm{Sb}}$}, where $r_\mrm{cb}$ is the radius of the convective boundary, $r_\mrm{Sb}$ the radius of the Schwarzschild boundary, and $H_{p,\mrm{Sb}}$ the pressure scale height at $r_\mrm{Sb}$. The four example simulations are 2.5D with $N_r = 512$. The time axis shows the time as a fraction of the thermal timescale $\tau_\mrm{th}(b)$. The dotted vertical line at $t = 0.3 \tau_\mrm{th}$ marks the end of the initial transient. Each data point corresponds to a time average over $0.05 \tau_\mrm{th}$.}
\label{fig:alpha_evolution}
\end{figure}

To better quantify when the penetration distance stops changing, we compute the radii $r_\mrm{Sb}$ and $r_\mrm{cb}$ of the Schwarzschild and convective boundaries, respectively, using simulation data averaged over the spherical angles and over time bins $0.05\, \tau_\mrm{th}$ long. We define $r_\mrm{cb}$ as the radius, where the drop in $\nabla(r) / \nabla_\mrm{ad}(r)$ is the steepest, see the inset in Fig.~\ref{fig:F71-nabla-evolution}. The dimensionless penetration distance is then $\alpha = (r_\mrm{cb} - r_\mrm{Sb})/H_{p,\mrm{Sb}}$, where the pressure scale height $H_{p,\mrm{Sb}}$ at the Schwarzschild boundary is derived from the same space- and time-averaged data. The time evolution of $\alpha$ is shown in Fig.~\ref{fig:alpha_evolution} for four example simulations. The statistical variation in the simulations with $b = 10^6$ is so large that all 20 data points seem to be consistent with a constant state. However, simulations with smaller boost factors cover many more convective timescales (see Table~\ref{tab:scales}) and the statistical variation is largely suppressed in the time averages. These simulations show a clear initial increase in $\alpha$ followed by a statistically constant state. Based on Fig.~\ref{fig:alpha_evolution}, we define the first $0.3 \tau_\mrm{th}$ to be an initial transient for the purpose of measuring $\alpha$. This transient is excluded from further analysis.

\begin{figure}
\includegraphics[width=\linewidth]{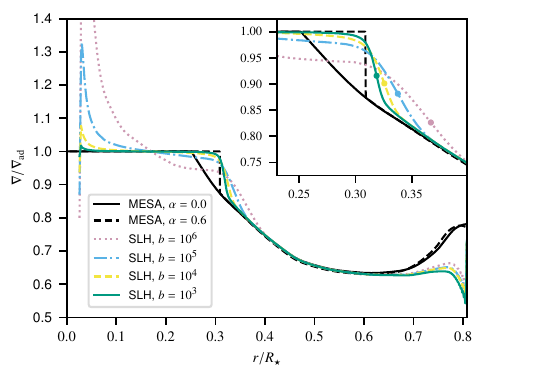}
\caption{Radial profiles of the temperature gradient $\nabla$ in 2.5D simulations with $N_r = 512$ and the boost factors given in the legend. The gradients are normalised using the adiabatic temperature gradient $\nabla_\mrm{ad}$. The curves are derived from simulation data averaged over the time interval $0.3 \tau_\mrm{th} \le t \le \tau_\mrm{th}$ except for the $b = 10^3$ simulation, for which we use the interval $0.3 \tau_\mrm{th} \le t \le 0.5\tau_\mrm{th}$. MESA models with $\alpha = 0.0$ (our initial condition) and $\alpha = 0.6$ are shown for comparison. The inset zooms in onto the convective-penetration layer with dots marking the location of the steepest gradient in $\nabla / \nabla_\mrm{ad}$.}
\label{fig:nabla-vs-b}
\end{figure}

We show the radial profiles of the temperature gradient $\nabla$ in 2.5D simulations with four different boost factors in Fig.~\ref{fig:nabla-vs-b}. The stratification in the core becomes increasingly closer to being adiabatic as the boost factor is decreased. However, the stratification is at least slightly sub-adiabatic at $r \gtrsim 0.17 R_\star$, which is well inside the Schwarzschild boundary at all four boost factors. The drop in the temperature gradient at the convective boundary (the outer boundary of the penetration layer) becomes increasingly steep with decreasing boost factor.

Our simulations suggest that a simplified model, often referred to as `step overshoot', which assumes that the penetration layer is fully mixed, perfectly adiabatic, and ends with a discontinuous jump to the radiative temperature gradient at the convective boundary is a good approximation to the thermal structure to be expected at the nominal luminosity ($b = 1$). A MESA model of a $15\,\mrm{M}_\odot$ ZAMS star computed using a parametric description of the penetration process is shown in Fig.~\ref{fig:nabla-vs-b} for comparison.\footnote{The `step overshoot' prescription that can be activated in MESA via an \texttt{inlist} file only mixes composition in the `overshoot' layer and the temperature gradient $\nabla$ is kept radiative. However, it is possible to enforce $\nabla = \nabla_\mrm{ad}$ and model convective penetration by adding a simple subroutine to \texttt{run\_star\_extras.f90} as we did in this example. Our MESA set-up is available on Zenodo (\url{https://doi.org/10.5281/zenodo.8127093}).} We use $\alpha = 0.6$ in this model, which is approximately the value we obtain by extrapolating the results of our 2.5D simulation to the actual stellar luminosity ($b = 1$) in Sect.~\ref{sec:luminosity-dependence}. In comparison, the popular alternative approach to model convective overshooting -- exponentially decaying diffusion of chemical elements -- assumes the temperature gradient to be radiative beyond the convective boundary. We do not show such a model in Fig.~\ref{fig:nabla-vs-b} because it would be indistinguishable from the model without any extra mixing at the convective boundary. It is possible if not likely that some slower form of mixing exists beyond the penetration layer. One could use the diffusive model beyond the layer covered by the adiabatic step-overshoot model, obtaining a combined model that \citet{anders2023a} refer to as `extended convective penetration.'

\subsection{Numerical convergence of the penetration distance}
\label{sec:numerical-convergence}

\begin{figure}
\includegraphics[width=\linewidth]{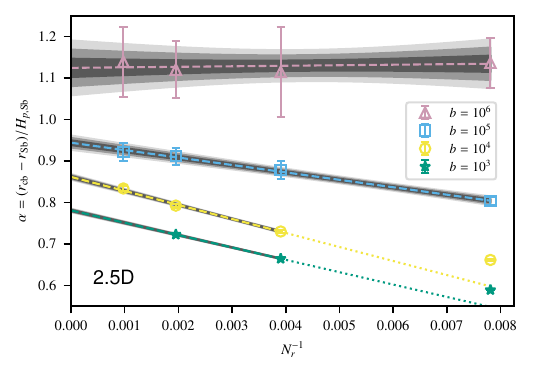}\\
\includegraphics[width=\linewidth]{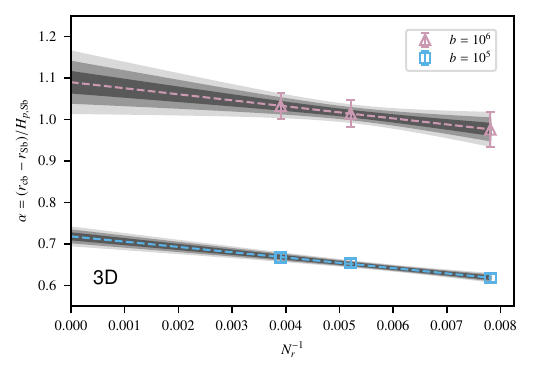}
\caption{Numerical convergence of the penetration distance $\alpha$ for 2.5D and 3D simulations with different boost factors $b$. The radial grid spacing is expressed as the inverse number of radial grid cells, $N_r^{-1}$, so that first-order convergence corresponds to a straight line. The error bars indicate $3\sigma$ intervals of statistical variation. The grey shading shows $\sigma$, $2\sigma$, and $3\sigma$ statistical-variation intervals of the fits. The dashed lines show the mean fit in each case and the dotted lines extrapolate the fits towards data points excluded from the fitting procedure.}
\label{fig:alpha_vs_resolution}
\end{figure}

For each boost factor and grid dimensionality, we run simulations on a range of grids in order to quantify and suppress discretisation effects. Figure~\ref{fig:alpha_vs_resolution} shows that 2.5D simulations with the highest boost factor ($b = 10^6$) are consistent with having the same value of $\alpha$ for all of our 2.5D grids ($N_r = 128$ to $N_r = 1024$), i.e. the simulations resolve all relevant scales well enough and $\alpha$ is numerically converged. The remaining sets of 2.5D and 3D simulations show some resolution dependence. For a second-order code, we would asymptotically expect the dependence $\alpha = \alpha_\infty + C N_r^{-2}$, where $\alpha_\infty$ is the unknown value of $\alpha$ for $N_r \to \infty$, and the constant $C$ sets the magnitude of the error term. We observe first-order convergence, $\alpha = \alpha_\infty + C N_r^{-1}$, instead, see Fig.~\ref{fig:alpha_vs_resolution}.

The diffusion length scales listed in Table~\ref{tab:scales} suggest that the smallest features one would expect to see in the entropy distribution around the core boundary are only marginally resolved on the finest of our grids in simulations with $b \in (10^3, 10^4)$, possibly explaining the slower-than-expected numerical convergence. However, there are other numerical effects that affect the mean thermal stratification in our simulations and possibly also the penetration distance. We include an a posteriori study of these effects in Appendix~\ref{sec:code_accuracy}. In Sect.~\ref{sec:test_kh2d}, we show that the core numerical solver of SLH is second-order accurate. However, our implementation of the constant-temperature boundary condition turns out to be only first-order accurate when there is a density gradient inside the computational domain, see Sect.~\ref{sec:test_diffusion} for details. Furthermore, we show in Sect.~\ref{sec:test_hot_bubble} that the effect of odd-even decoupling starts to dominate the remaining numerical errors when the physical solution becomes sufficiently well resolved. This effect, which is always present in numerical schemes on collocated grids, also reduces the convergence rate in our test problem. Both the inaccurate boundary condition and the odd-even-decoupling effect may have influenced the penetration distance via their influence on the mean thermal stratification, although it is difficult to judge which of them is more important.

Nevertheless, Fig.~\ref{fig:alpha_vs_resolution} shows that the penetration distance does converge and we can fit the first-order convergence law to the data. This way, we obtain a well-defined, extrapolated penetration distance $\alpha_\infty$ that would correspond to a hypothetical, infinitely fine grid and is not affected by the aforementioned problems restricting the convergence rate to first order. We employ the Monte Carlo procedure described in Sect.~\ref{sec:velocity-field}, so we obtain a statistical-variation range in addition to the mean fit. We exclude the simulations with  $N_r = 128$ ($N_r^{-1} = 7.8 \times 10^{-3}$) for the two lowest boost factors because these simulations seem to be severely under-resolved and they deviate significantly from the fits based on finer grids. None of the points included in the fits deviate from the fitting lines by more than $2\sigma$. The resulting values of $\alpha_\infty$ are summarised in Table~\ref{tab:alpha_infty}.

\begin{table}
\caption{Dimensionless penetration distance extrapolated to infinite grid resolution.}
\label{tab:alpha_infty}
\centering
\begin{tabular}{lll}
\toprule
DIM & $b$ & $\alpha_\infty$ \\
\midrule
2.5D & $10^6$ & $1.124 \pm 0.022$ \\
2.5D & $10^5$ & $0.944 \pm 0.006$ \\
2.5D & $10^4$ & $0.862 \pm 0.002$ \\
2.5D & $10^3$ & $0.781 \pm 0.001$ \\
\midrule
3D   & $10^6$ & $1.090 \pm 0.025$ \\
3D   & $10^5$ & $0.718 \pm 0.007$ \\
\bottomrule
\end{tabular}
\vspace{0.5em}
\tablefoot{The variables in the header are grid dimensionality DIM, boost factor $b$, and dimensionless penetration distance $\alpha_\infty$ with its $1\sigma$ range of statistical variation as extrapolated to infinite grid resolution in Sect.~\ref{sec:numerical-convergence}.}
\end{table}

\subsection{Luminosity dependence of the penetration distance}
\label{sec:luminosity-dependence}

\begin{figure}
\includegraphics[width=\linewidth]{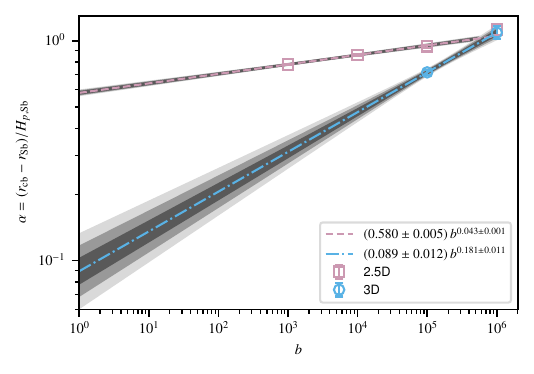}
\caption{Penetration distance $\alpha$ extrapolated to an infinitely fine grid (see Fig.~\ref{fig:alpha_vs_resolution}) as a function of the boost factor $b$. Assumed power-law dependencies $\alpha(b)$ are fitted to the 2.5D and 3D data points separately. The grey shading shows $\sigma$, $2\sigma$, and $3\sigma$ statistical-variation intervals of the fits.}
\label{fig:alpha_vs_b_1}
\end{figure}

Having suppressed discretisation effects, we are left with a single estimate of the penetration distance and its range of statistical variation for each boost factor and grid dimensionality. We show these data points in Fig.~\ref{fig:alpha_vs_b_1}. The 2.5D data points suggest a weak, although highly statistically significant, dependence of $\alpha$ on the boost factor $b$. We first blindly fit power laws to the 2.5D and 3D data sets individually using the Monte Carlo procedure described in Sect.~\ref{sec:velocity-field}. In the 2.5D case, the scaling exponent is only $0.043 \pm 0.001$ and the penetration distance extrapolated to the nominal luminosity ($b = 1$) is $\alpha_{2.5D} = 0.580 \pm 0.005$. If we assume that the $\alpha(b)$ dependence for 3D simulations can also be described by a single power law, we obtain a much steeper scaling with an exponent of $0.181 \pm 0.011$. After extrapolating over five orders of magnitude, the power law would then imply a penetration distance of only $\alpha_{3D} = 0.089 \pm 0.012$. However, our current set of 3D simulations on its own does not contain enough evidence for the single-power-law model.

\begin{figure}
\includegraphics[width=\linewidth]{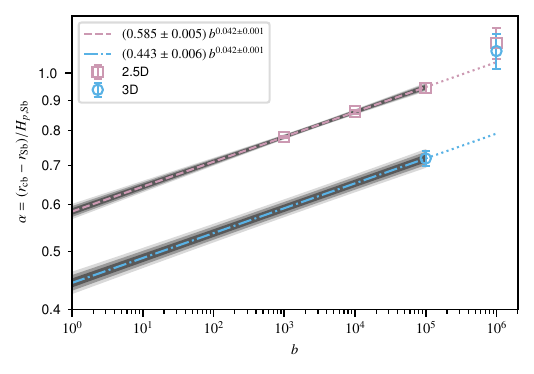}
\caption{As in Fig.~\ref{fig:alpha_vs_b_1}, but excluding the simulations with $b = 10^6$, fitting a power law to the remaining 2.5D data points and, assuming that the scaling exponent is the same in 3D, shifting the power law to pass through the data point based on 3D simulations with $b = 10^5$. The statistical variation associated with that data point is added to that associated with the scaling law.}
\label{fig:alpha_vs_b_2}
\end{figure}

It is also possible that there is a power-law dependence $\alpha(b)$ but only for sufficiently low boost factors $b$, e.g.\ due to compressibility effects.\footnote{The Mach number $\mach$ sometimes exceeds $0.3$ in simulations with \mbox{$b = 10^6$}. Ram pressure fluctuations are of the order of $\mach^2$.} Indeed, the 2.5D data point with $b = 10^6$ deviates from the scaling law discussed above and shown in Fig.~\ref{fig:alpha_vs_b_1} by $3.2\sigma$. Therefore, we now exclude the $b = 10^6$ data points and explore the alternative assumption that the power law's slope is the same for 2.5D and 3D simulations with $b \le 10^5$. Figure~\ref{fig:alpha_vs_b_2} shows that the scaling exponent based on the three remaining 2.5D data points remains essentially unchanged ($0.042 \pm 0.001$) because the statistical weight of the $b = 10^6$ data point considered in the fit in Fig.~\ref{fig:alpha_vs_b_1} is low. The largest deviation from the power law is only $0.5\sigma$ now. The penetration distance implied by the 2.5D simulations at the nominal luminosity does not change significantly either ($\alpha_{2.5D} = 0.585 \pm 0.005$). However, the much shallower scaling implies a much larger extrapolated penetration distance of $\alpha_{3D} = 0.443 \pm 0.006$ for the 3D simulations if we fix the scaling exponent and simply shift the 2.5D-based power law to make it pass through the 3D-based data point at $b = 10^5$.

\section{Summary and discussion}
\label{sec:summary-discussion}

Observational evidence strongly suggests that the convective cores of intermediate-mass and massive, main sequence stars are substantially larger that what is predicted by linear stability analysis. The process of convective penetration, first described by \citet{roxburgh1978a}, seems to be able to enlarge the convective cores of main sequence stars enough to reconcile them with observations. However, the penetration distance depends on the complex physics of 3D turbulent convection in the core and such effects cannot be consistently captured in analytic treatments.

We performed a large set of simulations of convective penetration in a chemically homogeneous model of a $15\,\mrm{M}_\odot$ ZAMS star. We make it possible to cover the thermal timescale by increasing the luminosity and decreasing the opacity by the same large boost factor $b$ ranging from $10^3$ to $10^6$. This scaling procedure preserves the value of the radiative temperature gradient. We run both 2.5D and 3D simulations using our time-implicit, low-Mach-number code SLH and quantify their numerical convergence using a range of computational grids.

We argue that mass entrainment into the convective core cannot stop in the adiabatic approximation and show an adiabatic simulation that demonstrates this. Indeed, the star becomes fully convective in the adiabatic simulation. On the other hand, simulations with radiative diffusion reach a statistically stationary state with a convective core and radiative envelope on the thermal timescale.

All of our simulations with radiative diffusion show a clear penetration layer, the thickness of which becomes statistically constant in time once thermal equilibrium has been reached. As expected, simulations performed with large luminosity boost factors show large deviations from adiabacity in the core. However, the core, including the bulk of the penetration layer, becomes nearly adiabatic as the luminosity is decreased. This result qualitatively agrees with the simulations of penetrative convection in similar stellar environments performed by \citet{baraffe2023a}, \citet{mao2023a}, and \citet{blouin2023a},  as well as with the much more idealised simulations of \citet{anders2022a}.

There is a steep jump from the adiabatic to the radiative temperature gradient at the outer boundary of the penetration layer. This means that the popular simplified model, often referred to as `step overshoot', which assumes that the penetration layer is fully mixed, perfectly adiabatic, and ends with a discontinuous jump to the radiative temperature gradient at the convective boundary, is a good approximation of the thermal structure to be expected at the nominal luminosity ($b = 1$).

We find strong evidence for a weak dependence of the dimensionless penetration distance $\alpha$ on the boost factor $b$. Weak if any dependence of the core size on $b$ is expected also from the theoretical point of view. In the Roxburgh criterion (Eq.~\eqref{eq:roxburgh_criterion}), the energy flux $\overline{\Gamma}$ due to nuclear sources scales in proportion to $b$; as does the radiative flux $\overline{F}$ because we scale the opacity in inverse proportion to $b$. The turbulent dissipation rate $\Phi_0$ is expected to scale with $u_\mrm{conv}^3/L_\mrm{turb}$, where $L_\mrm{turb}$ is an integral length scale of the turbulence. If $L_\mrm{turb}$ is fixed and we consider adiabatic convection, we have the usual scaling $u^3 \propto b$, and so both sides of Eq.~\eqref{eq:roxburgh_criterion} scale in proportion to $b$ and the size of the core cannot depend on $b$. However, we show in Sect.~\ref{sec:velocity-field} and Fig.~\ref{fig:u_conv_vs_b} that the velocity scaling in our non-adiabatic simulations is slightly shallower than that of adiabatic convection. This changes the relative importance of the dissipation term as $b$ is varied over several orders of magnitude. The slight change in the core radius $r_\mrm{cb}$ itself may affect the dissipation rate given that it is reasonable to assume that $L_\mrm{turb} \propto r_\mrm{cb}$. We must also keep in mind that \citet{roxburgh1989a} assumes the velocity vector to vanish at the convective boundary and that contributions from terms quadratic in fluctuations can be neglected. Some of those terms may not be negligible in our simulations, especially because some energy flux is needed to drive the waves we clearly see in the radiative envelope.

Extrapolating the results of our 2.5D simulations to $b = 1$, we derive a rather large penetration distance of $\alpha = 0.585 \pm 0.005$. Our set of 3D simulations covers only two relatively large boost factors at the moment. For this reason, we show two possible extrapolations to $b = 1$ in Sect.~\ref{sec:luminosity-dependence}, which give penetration distances $\alpha = 0.089 \pm 0.012$ and $\alpha = 0.443 \pm 0.006$. These two values differ by a factor of five, but both are broadly consistent with observations. For stars of similar masses, the compilation of \citet[][their Fig.~12]{anders2023a} of observational constraints on the penetration distance from asteroseismology shows measurements and upper limits that roughly span our two $\alpha$ estimates based on  3D simulations. \citet{brott2011a} discuss a drop in rotational velocity observed at a certain value of surface gravity in a sample of stars in the Large Magellanic Cloud. Interpreting the drop as the terminal-age main sequence, their calibration gives $\alpha = 0.34 \pm 0.1$ for stellar models in the mass range from approximately $10$ to $20\,\mrm{M}_\odot$. The calibration of \citet{claret2016a} based on eclipsing binaries shows that $\alpha$ increases from zero at $1.2\,\mrm{M}_\odot$ to ${\approx}\,0.2$ at $2.1\,\mrm{M}_\odot$, where it stops changing and remains approximately the same up to $4.4\,\mrm{M}_\odot$. The fitting formula given by Eqs.~(11)\,--\,(16) of \citet{jermyn2022a}, which is based on the simulations and theory of \citet{anders2022a}, gives $\alpha = 0.19$ for a $15\,\mrm{M}_\odot$ star.

The results presented here demonstrate only the most basic properties of the rich data set we have created. An analysis based on Reynolds averaging is already in progress and is expected to quantify differences between our 2.5D and 3D simulations and become a basis for constructing simplified models of the penetration process. We are also working on extending our simulations ---especially those in 3D geometry--- to lower luminosity boost factors. Although the simulations are based on a realistic stellar model, the model is chemically homogeneous and does not rotate or contain magnetic fields. Additionally, our computational grid currently covers a spherical wedge rather than the full sphere. All of these assumptions and simplifications should be gradually relaxed in order to generate predictions for a wide range of observed stars.

\begin{acknowledgements}
We are grateful to Raphael Hirschi and his group at Keele University as well as to Saskia Hekker at the Heidelberg Institute for Theoretical Studies for fruitful discussions and comments. We acknowledge support by the Klaus Tschira Foundation. This work is funded by the Deutsche Forschungsgemeinschaft (DFG, German Research Foundation) under Germany's Excellence Strategy EXC 2181/1 - 390900948 (the Heidelberg STRUCTURES Excellence Cluster). This work has received funding from the European Research Council (ERC) under the European Union’s Horizon 2020 research and innovation programme (Grant agreement No.\ 945806). PVFE was supported by the U.S. Department of Energy through the Los Alamos National Laboratory (LANL). LANL is operated by Triad National Security, LLC, for the National Nuclear Security Administration of the U.S. Department of Energy (Contract No. 89233218CNA000001). This work has been assigned a document release number LA-UR-23-26456.
\end{acknowledgements}

\bibliographystyle{aa}
\bibliography{convective_penetration}

\begin{appendix}
\section{2.5D convection without mass entrainment}
\label{sec:core_convection_experiment}

To judge the importance of mass entrainment on the convective  velocity in 2.5D convection, we perform a simple numerical experiment: we run 2.5D simulations of core convection using the same setup as in the convective-penetration simulations except that we impose a reflective boundary condition in the outer parts of the initially isentropic core. This way, we eliminate any mass entrainment from the radiative envelope. The volume heating and opacity profile are the same as those in the penetration simulations with the boost factor $b =10^4$. We keep the temperature at the outer radial boundary fixed at its initial value. The outer boundary is at $r = 8.5003125 \times 10^{10}$\,cm, so that we can achieve exactly the same grid spacing as in the simulations with the radiative envelope. We stop the set of core-only simulations after $0.3 \tau_\mrm{th}(b)$.

The elimination of the radiative envelope and of any mass entrainment leads to major qualitative and quantitative changes in the velocity field. The simulations with the radiative envelope and $b = 10^4$ contain a large number of vortices of different sizes, the distribution and shapes of which change chaotically in time (see Fig.~\ref{fig:velocity-distribution-2.5D} for a representative snapshot). On the other hand, the core-only simulations develop a single, smooth convective cell filling the core early on, which later splits into two large cells. The relative sizes of those cells change on short timescales but the cells remain very smooth with essentially no small-scale structure.

\begin{figure}
\includegraphics[width=\linewidth]{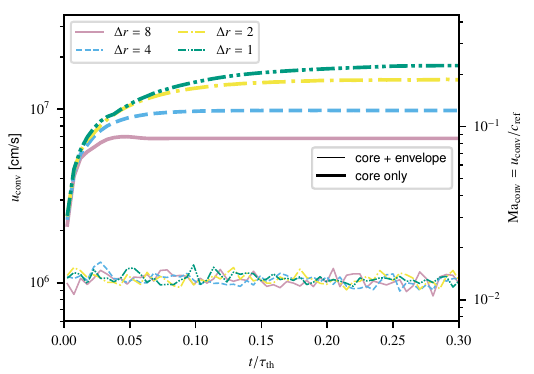}
\caption{Time evolution of the convective velocity $u_\mrm{conv}$ in two series of 2.5D simulations with radiative diffusion: the simulations of convective penetration with mass entrainment from the radiative envelope (`core + envelope') and similar simulations without the radiative envelope (`core only'). All of the simulations have the boost factor $b = 10^4$. The time axis shows simulation time as a fraction of the thermal timescale $\tau_\mrm{th}(b)$. Each data point corresponds to a time average over $0.005 \tau_\mrm{th}$. The right vertical axis shows the corresponding Mach numbers computed using the reference sound speed $c_\mrm{ref} = 7.96 \times 10^7$\,cm\,s$^{-1}$. Grid resolution is given in terms of the dimensionless radial grid spacing $\Delta r$ such that a grid with $N_r = 1024$ in the `core + envelope' case has $\Delta r = 1$.}
\label{fig:u_conv_evolution_appendix}
\end{figure}

Figure~\ref{fig:u_conv_evolution_appendix} compares the time evolution of the convective velocity in the two sets of simulations. The velocity fluctuates around the same, nearly constant, function in all four simulations with the radiative envelope, independently of their grid resolution. On the other hand, we see a fast increase in the convective velocity in the core-only simulations. It ultimately reaches a constant value (after averaging out rapid changes as done in Fig.~\ref{fig:u_conv_evolution_appendix}), which increases upon grid refinement. Starting with the coarsest grid, the velocity on the next finer grid is larger by factors of $1.45$, $1.51$, and $1.21$. The last ratio cannot be interpreted as a sign of numerical convergence because the convective velocity was still increasing in the highest-resolution simulation when it was stopped and it also showed much less variability on short timescales (not visible in Fig.~\ref{fig:u_conv_evolution_appendix}) than the lower-resolution simulations. The velocities in all of the core-only simulations are much larger than those in the analogous simulations of convective penetration.

These effects can be understood as follows. Because the core-only simulations lack a stable layer on top of the convective core, they cannot develop small-scale shear instabilities at that interface, leading to much more mundane convective flows. There is also no mass entrainment of high-entropy material from the stable envelope and therefore no buoyancy braking associated with this phenomenon. The only remaining mechanism dissipating kinetic energy is numerical diffusion, the magnitude of which decreases upon grid refinement. Because the physical scale of the convective cells is always the same, the timescale of reaching a balance between the driving of convection and numerical dissipation becomes longer upon grid refinement and the flows accumulate more kinetic energy by the time a stationary state is reached. The fact that radiative diffusion decreases temperature and density fluctuations, hence also the buoyancy driving the convection, does not seem to be sufficient to obtain low-viscosity flows with a well-defined velocity in two spatial dimensions.

\section{Accuracy of the SLH code}
\label{sec:code_accuracy}

Motivated by the first-order convergence of the penetration distance $\alpha$ upon grid refinement (see Sect.~\ref{sec:numerical-convergence} and Fig.~\ref{fig:alpha_vs_resolution}), we re-evaluated the accuracy of the SLH code on a set of three test problems designed to isolate different aspects of the fluid dynamics simulations reported in this work. All three test problems are 2D and formulated such that one can obtain a `converged' solution on a sufficiently fine grid and use it as a reference to quantify numerical errors affecting solutions on grids coarser than the reference grid. More specifically, we define $L_1$ errors in the distribution of an arbitrary quantity $q$ to be
\begin{align}
    L_1 = \frac{1}{\sigma^\mrm{ref}} \frac{\sum_{i=1}^{N_x} \sum_{j=1}^{N_y} \left| q_{i,j} - q_{i,j}^\mrm{ref} \right|}{N_x N_y},
\end{align}
where $q_{i,j}$ is the solution on a given grid and $q_{i,j}^\mrm{ref}$ is the reference solution binned onto the same grid of $N_x \times N_y$ cells.\footnote{We perform the binning using mass-weighted averaging, which is appropriate to the quantities used in our tests (specific internal energy and specific entropy).} The $L_1$ error is normalised using the standard deviation $\sigma^\mrm{ref}$ of $q$ in the reference solution because the fluctuations of interest are often much smaller than the mean value.  The spatial and temporal discretisation in the same as in the simulations of convective penetration, see Sect.~\ref{sec:methods} for details.

In the first test problem, a Kelvin-Helmoltz instability with radiative diffusion (Sect.~\ref{sec:test_kh2d}), we exclude gravity and use periodic boundary conditions to eliminate any boundary effects. We demonstrate that the numerical scheme is second-order accurate in this case. In the second test problem, we eliminate hydrodynamics and solve a diffusion problem with constant-temperature boundary conditions (Sect.~\ref{sec:test_diffusion}). This problem reveals a subtle effect that makes our original implementation of constant-temperature boundaries first-order accurate when there is a density stratification inside the computational domain. We fix this issue and show that the new implementation, to be used in future simulations, is second-order accurate. Finally, we solve a problem involving the buoyant rise of a `hot bubble' under the influence of radiative diffusion in the isentropic core of our $15\,\mrm{M}_\odot$ stellar model. This last problem demonstrates that the improved constant-temperature boundary condition is second-order accurate also when combined with hydrodynamics. However, we find that errors related to the infamous odd-even-decoupling effect limit the convergence rate to first order once relative errors have dropped below $\approx 10^{-2}$. As discussed in Sect.~\ref{sec:test_hot_bubble}, this effect is intrinsic to collocated numerical schemes and it can only be suppressed, not eliminated.

\subsection{Kelvin-Helmholtz instability}
\label{sec:test_kh2d}

\begin{figure}
\includegraphics[width=\linewidth]{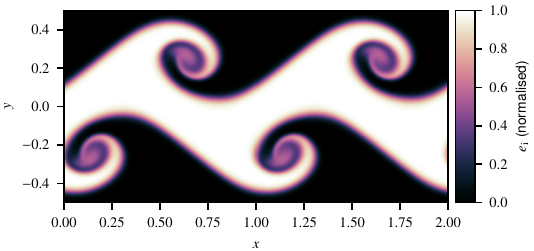}
\caption{Reference solution to the 2D Kelvin-Helmholtz problem in terms of the specific internal energy $e_\mathrm{i}$ normalised to range from 0 to 1. The solution was computed on a $4096 \times 2048$ grid.}
\label{fig:kh2d_reference_solution}
\end{figure}

The initial condition for this test problem is
\begin{align}
\rho &= \gamma \left[ 1.25 - 0.5 \eta(y) \right], \\
u &= 0.01 \left[ 1 - 2\eta(y)\right], \\
v &= 0.001 \sin(2 \pi x), \\
p &= 1,
\end{align}
where $\rho$ is the density, $u$ and $v$ are the components of the velocity vector, $p$ is the pressure, and the function
\begin{align}
    \eta(y) = 
    \begin{cases}
      \frac{1}{2} \big\{ 1+\sin \left[ 16\pi(y+0.25)\right] \big\}, &\text{for}\ y > -\frac{9}{32}\ \text{and}\ y < -\frac{7}{32}, \\
      1, &\text{for}\ y \ge -\frac{7}{32}\ \text{and}\ y \le \frac{7}{32}, \\
      \frac{1}{2}\big\{1-\sin\left[16\pi(y-0.25) \right] \big\}, &\text{for}\ y > \frac{7}{32}\ \text{and}\ y < \frac{9}{32}, \\
      0,  &\text{otherwise}
    \end{cases}
\end{align}
defines a smooth transition between the shearing layers. The smoothness of the transition makes it possible to obtain numerically converged solutions. The equation of state is that of an ideal monatomic gas with the ratio of specific heats $\gamma = 1.4$ and mean molecular weight $\mu = 1$. The initial Mach number is approximately $0.01$. We include radiative diffusion with a constant conductivity coefficient of $\chi = 3.61 \times 10^{-5}\,\mathcal{R}$, where $\mathcal{R}$ is the gas constant per unit of mass. We stop the simulations at $t = 80$ when the instability has evolved well into its non-linear phase, see Fig.~\ref{fig:kh2d_reference_solution}. Our choice of $\chi$ is such that the diffusion affects the instability in a significant way without suppressing it completely. In the absence of hydrodynamics, the radiative diffusion on its own would widen the smooth transitions between the shear layers approximately by a factor of three by $t = 80$. Depending on whether we take the wavelength of the initial perturbation (unity) or the initial width of the transitions between the shear layers ($\frac{1}{16}$) as a reference length scale, the P\'eclet number of the flow is ${\approx}\,870$ and ${\approx}\,55$, respectively.

\begin{figure}
\includegraphics[width=\linewidth]{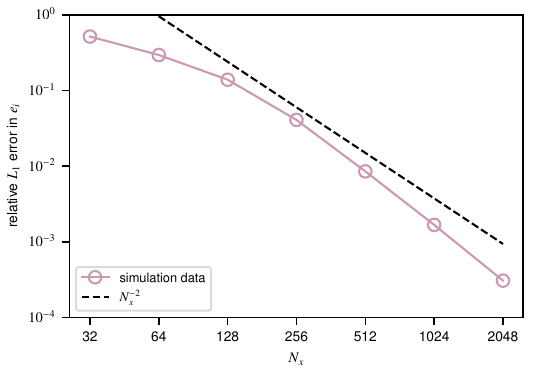}
\caption{Numerical convergence of the solutions to the Kelvin-Helmholtz problem. Relative $L_1$ errors in the specific internal energy $e_i$ are shown as a function of the grid resolution $N_x$ along the $x$ axis. A second-order scaling is shown for reference.}
\label{fig:kh2d_L1}
\end{figure}

We compute solutions to this test problem on equidistant, Cartesian grids ranging from $32 \times 16$ to $4096 \times 2048$ cells. The latter serves as the reference solution. The $L_1$ errors in the specific internal energy $e_i$ are shown in Fig.~\ref{fig:kh2d_L1}. As expected, the numerical convergence is slow on very coarse grids, which do not resolve important features of the solution. Convergence slightly faster than second-order is obtained on grids with at least $256$ cells along the $x$ axis. Our use of parabolic reconstruction, which is formally third-order accurate, is the most likely reason for the fast convergence.

\subsection{Heat diffusion with constant-temperature boundaries}
\label{sec:test_diffusion}

\begin{figure}
\includegraphics[width=\linewidth]{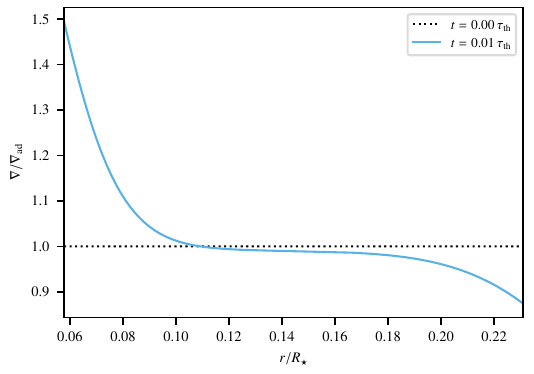}
\caption{Angle-averaged reference solution to the diffusion problem computed on a 2.5D grid of $2048 \times 512$ cells and visualised in terms of the temperature gradient. The isentropic initial state has $\nabla / \nabla_\mrm{ad} = 1$.}
\label{fig:diffusion_reference_solution}
\end{figure}

To test our constant-temperature boundary conditions, we set up a problem that involves diffusion but no hydrodynamics. We use part of the initially isentropic core of our $15\,\mrm{M}_\odot$ stellar model (Sect.~\ref{sec:stellar-model-simplifications}), namely the radial range from $2 \times 10^{10}$\,cm to $8 \times 10^{10}$\,cm, as an initial condition. We fix the temperatures at both radial boundaries to their initial values, turn off volume heating, and include radiative diffusion with the original stellar opacity profile. We use $100$ long, implicit time steps to reach $1\%$ of the thermal timescale\footnote{This is the thermal timescale of the whole stratification as included in the simulations of convective penetration. We use the same value in these simulations for simplicity.} ($589$\,yr) and we stop the simulations at this point. The resulting change in the temperature gradient is shown in Fig.~\ref{fig:diffusion_reference_solution}.

\begin{figure}
\includegraphics[width=\linewidth]{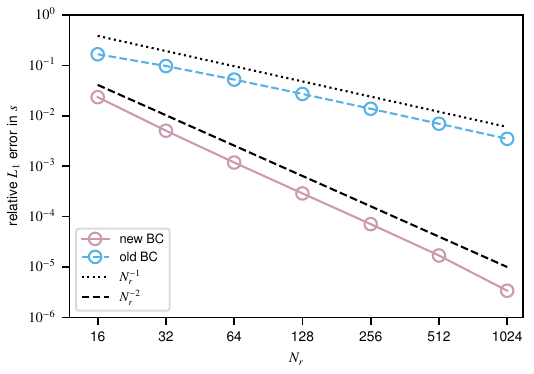}
\caption{Numerical convergence of the solutions to the diffusion problem with the constant-temperature boundary condition used in the simulations of convective penetration (old BC) and with an improved version of the same boundary condition (new BC). Relative $L_1$ errors in the specific entropy $s$ are shown as a function of the radial grid resolution $N_r$. First- and second-order scaling laws are shown for reference.}
\label{fig:diffusion_L1}
\end{figure}

Although the stratification is spherically symmetric, we use 2.5D grids covering polar angles in the range $82.5^\circ \le \vartheta \le 97.5^\circ$. We solve this problem on grids ranging from $16 \times 4$ to $2048 \times 512$ cells. The latter serves as the reference solution. The $L_1$ errors in the specific internal entropy $s$ are shown in Fig.~\ref{fig:diffusion_L1}. The constant-temperature boundary condition as implemented for the simulations of convective penetration reported in this work is denoted `old BC' in the figure. Clearly, the solutions converge at first order only. Re-evaluating the implementation, we identified the following subtle issue that causes the slow convergence. When computing the diffusive flux at a given cell face, we use the states in the two cells around the interface to compute estimates of the temperature gradient, opacity (which is fixed), and the $T^3/\rho$ factor that enters the thermal conductivity at the cell interface. These approximations are second-order accurate for smooth solutions. At the domain boundary, we fix the temperature in the first ghost cell and use reflective boundary conditions for the density and momentum as described in Sect.~\ref{sec:hydro-simulations}. The temperature varies smoothly across the domain boundary, so the estimate of the temperature at the interface is second-order accurate. However, if there is a density stratification inside the computational domain the reflective boundary condition forces the component of the density gradient normal to the boundary to suddenly vanish at the domain boundary, introducing a first-order error term and explaining the first-order convergence observed in Fig.~\ref{fig:diffusion_L1}. Because our simulations of convective penetration are long enough to form a statistically stationary state, first-order errors would have propagated from the outer domain boundary throughout the simulation domain, possibly reducing the numerical convergence rate of the mean thermal stratification and of the penetration distance.

We improved our implementation of the constant-temperature boundary condition such that the state at the interface is obtained by extrapolating the state from the physical domain using second-order-accurate expressions. We also compute the temperature gradient using one-sided, second-order-accurate finite differences, fixing the temperature exactly at the domain boundary (rather than in the first ghost cell). The same test problem converges at second order with the new implementation, see Fig.~\ref{fig:diffusion_L1} (the `new BC' curve). The improved implementation will be used in future simulations.

\subsection{Buoyant rise of a hot bubble}
\label{sec:test_hot_bubble}

Our third and most challenging test problem involves the buoyant rise of a single `hot bubble' in the initially isentropic core of our $15\,\mrm{M}_\odot$ stellar model (Sect.~\ref{sec:stellar-model-simplifications}) under the influence of radiative diffusion with constant-temperature boundary conditions. The 2.5D simulation domain spans the radial range from $2 \times 10^{10}$\,cm to $8 \times 10^{10}$\,cm and polar angles $60^\circ \le \vartheta \le 120^\circ$. The `hot bubble' is modelled as an entropy perturbation
\begin{align}
    s = s_0 + 
    \begin{cases}
        \Delta s \cos\left( \frac{\pi}{2} \frac{\zeta}{r_\mrm{b}} \right)\ \mrm{for}\ \zeta < r_\mrm{b},\\
        0\ \mrm{otherwise},
    \end{cases}
\end{align}
where $s_0$ is the specific entropy of the background stratification, $\Delta s$ the amplitude of the perturbation, $r_\mrm{b} = 10^{10}$\,cm the radius of the bubble, and $\zeta$ the distance from the centre of the bubble located at $r = 3.5 \times 10^{10}$\,cm and $\vartheta = 0^\circ$. We do not include volume heating in the test problem. Radiative diffusion is included with the opacity profile identical to that used in our simulations of convective penetration with the boost factors of $b \in (10^3, 10^4, 10^5)$. The perturbation amplitude $\Delta s$ is $10^5$, $4 \times 10^5$, and $1.6 \times 10^6$\,erg\,g$^{-1}$\,K$^{-1}$ and the simulations are stopped after $1.6 \times 10^5$, $8 \times 10^4$, and $4 \times 10^4$\,s with boost factors of $10^3$, $10^4$, and $10^5$, respectively. We keep the temperatures at both radial boundaries at their initial values using both implementations of the constant-temperature boundary condition discussed in Sect.~\ref{sec:test_diffusion}.

\begin{figure*}
\includegraphics[width=\linewidth]{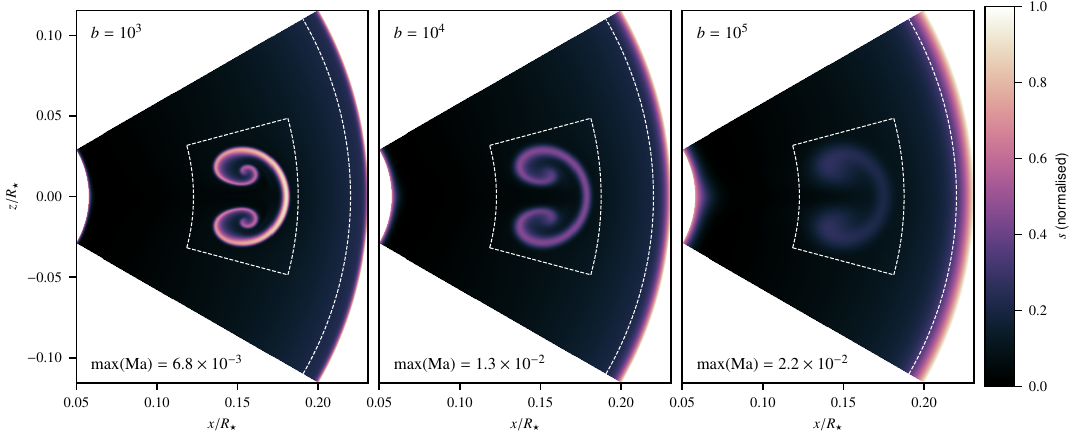}
\caption{Reference solutions to the hot-bubble problem with three different boost factors $b$ in terms of the specific entropy $s$ normalised to range from $0$ to $1$. The solutions were computed a 2.5D grid of $2048 \times 2048$ cells. The `bubble' and `boundary' subdomains, in which we compute the $L_1$ errors shown in Fig.~\ref{fig:hot_bubble_L1}, are marked using the white dashes lines.}
\label{fig:hot_bubble_reference_solutions}
\end{figure*}

Buoyancy makes the bubble accelerate upwards. The central part of the bubble, where the entropy perturbation is largest, gradually overtakes the outer parts and the bubble turns into a pair of vortices. Radiative diffusion decreases the entropy contrast between the bubble and its surroundings. It also generates a mean entropy gradient in the background stratification with prominent boundary layers at the radial domain boundaries. We stop the simulations at a stage when the bubble has become strongly deformed but while it is still possible to obtain a numerically converged solution. We run this test problem on grids ranging from $16^2$ to $2048^2$ cells. The latter serves as the reference solution. The distributions of specific entropy at the end of the reference simulations are shown in Fig.~\ref{fig:hot_bubble_reference_solutions}. The figure also includes the corresponding maximum Mach numbers reached in the simulation domain, which (by our choise of $\Delta s$) closely match the convective velocity in our simulations of convective penetration at the same boost factors, see Fig.~\ref{fig:u_conv_vs_b}.

\begin{figure}
\includegraphics[width=\linewidth]{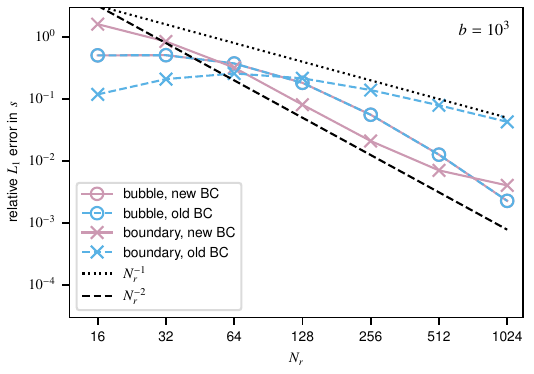}\\
\includegraphics[width=\linewidth]{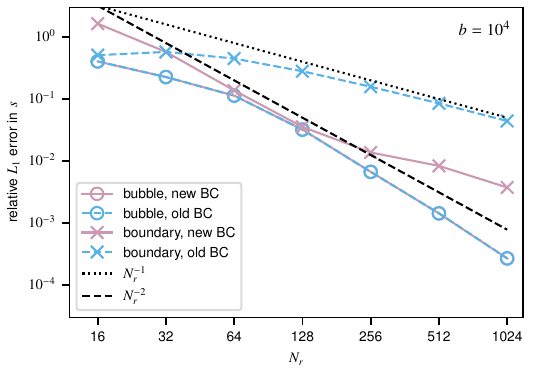}\\
\includegraphics[width=\linewidth]{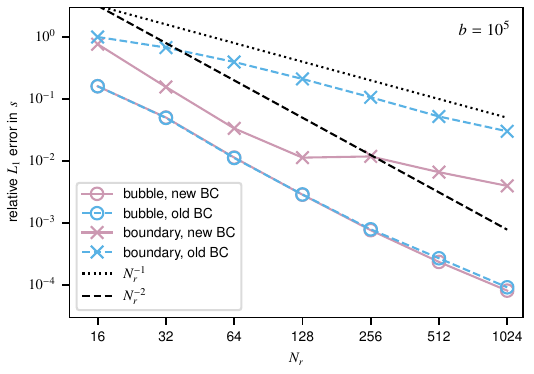}
\caption{Numerical convergence of the solutions to the hot-bubble problem with the constant-temperature boundary condition used in the simulations of convective penetration (old BC) and with an improved version of the same boundary condition (new BC). Relative $L_1$ errors in the specific entropy $s$ are shown as a function of the radial grid resolution $N_r$ for three boost factors $b$. First- and second-order scaling laws are shown for reference.}
\label{fig:hot_bubble_L1}
\end{figure}

We compute $L_1$ errors in the specific entropy $s$ in two subdomains, which we call `bubble' and `boundary' as depicted in Fig.~\ref{fig:hot_bubble_reference_solutions}. The $L_1$ errors for the two subdomains, three boost factors $b$, and two implementations of the constant-temperature boundary condition are shown in Fig.~\ref{fig:hot_bubble_L1}. As expected, the convergence is slow on coarse grids and low boost factors, i.e. when important features of the solutions are unresolved. We now focus on finer grids ($N_r \gtrsim 64$). The $L_1$ errors for the `bubble' subdomain are essentially the same with both kinds of boundary conditions and second-order convergence is obtained on sufficiently fine grids. The only exception is a slight decrease in the convergence rate on the finest grids in simulations with $b = 10^5$, which may be caused by the influence of the boundary effect discussed below. Indeed, the improved constant-temperature boundary condition, detailed in Sect.~\ref{sec:test_diffusion}, gives slightly smaller errors for the `bubble' subdomain with $N_r \gtrsim 512$.

\begin{figure}
\includegraphics[width=\linewidth]{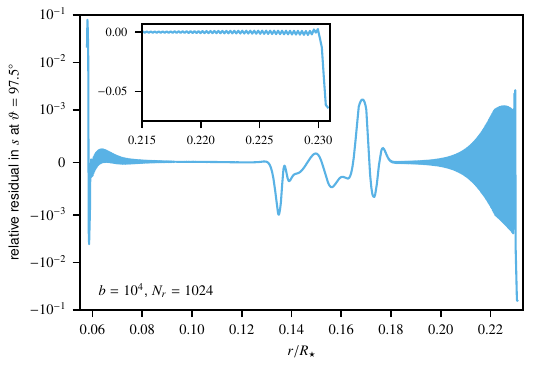}
\caption{Relative residuals in the specific entropy $s$ along the radial ray at $\vartheta = 97.5^\circ$ in a simulation of the hot-bubble problem with $b = 10^4$ computed on a $1024 \times 1024$ grid. The inset shows the outer boundary layer on a linear scale, which makes it easier to judge the magnitude of the odd-even-decoupling effect.}
\label{fig:odd_even_decoupling}
\end{figure}

The new boundary condition greatly improves the convergence rate in the boundary layer, in agreement with the results of the diffusion test reported in Sect.~\ref{sec:test_diffusion}. However, the convergence rate drops to (approximately) first order once the relative errors have dropped below ${\approx}\,10^{-2}$, a threshold almost independent of the boost factor. The reason for this behaviour is the infamous effect of odd-even decoupling that is unavoidable in numerical schemes working with collocated grids. The decoupling occurs when the mean stratification changes and it takes the form of grid-scale oscillations in the radial direction. We visualise this effect in Fig.~\ref{fig:odd_even_decoupling} using the signed residuals in $s$ along the radial ray at $\vartheta = 97.5^\circ$, which passes through one of the vortices, in the simulation with $b = 10^4$ computed on a $1024 \times 1024$ grid. Whereas the residuals are smooth in the radial range occupied by the bubble (c.f.\ Fig.~\ref{fig:hot_bubble_reference_solutions}), they tend to oscillate between positive and negative values close to the radial domain boundaries.

Similar oscillations occur close to the outer radial boundary in our simulations of convective penetration, possibly influencing the numerical convergence rate of the mean thermal stratification and of the penetration distance. The decoupling effect can be suppressed by increasing numerical diffusivity (via the pressure-diffusion term in the AUSM+-up flux function, see \citet{liou2006a}). However, doing so only trades decoupling errors for diffusive errors, reducing the effective resolving power of the numerical scheme.

\end{appendix}
\end{document}